\documentclass[letterpaper]{article} 
\usepackage{aaai2026}  
\nocopyright
\usepackage{times}  
\usepackage{helvet}  
\usepackage{courier}  
\usepackage[hyphens]{url}  
\usepackage{graphicx} 
\usepackage{natbib}  
\usepackage{caption,subcaption,enumitem, booktabs,amsmath} 
\usepackage{amssymb, amsthm, mathtools}
\usepackage{algorithm}
\usepackage{algorithmic}
 \usepackage{multirow}
\usepackage{newfloat}
\usepackage{listings}
\DeclareCaptionStyle{ruled}{labelfont=normalfont,labelsep=colon,strut=off} 
\floatstyle{ruled}
\newfloat{listing}{tb}{lst}{}
\floatname{listing}{Listing}
\newtheorem{definition}{Definition}

\usepackage{xcolor}

\newif\ifanon
 \anonfalse       

\newcommand{\redacted}{\textcolor{gray}{[Redacted]}}
\newcommand{\ifanonelse}[2]{\ifanon #1\else #2\fi}

\title{Beyond Polarization: Opinion Mixing and Social Influence in Deliberation}

\author {
    Mohak Goyal\textsuperscript{\rm 1},
    Lodewijk Gelauff\textsuperscript{\rm 2, \rm 3},
    Naman Gupta\textsuperscript{\rm 1},
    Ashish Goel\textsuperscript{\rm 1},
    Kamesh Munagala\textsuperscript{\rm 4}
}
\affiliations {
    \textsuperscript{\rm 1}Department of Management Science \& Engineering, Stanford University, USA\\
    \textsuperscript{\rm 2}Center on Democracy, Development, and the Rule of Law, Stanford University, USA\\
    \textsuperscript{\rm 3}Generation Lab, Washington D.C., USA\\
    \textsuperscript{\rm 4}Department of Computer Science, Duke University, USA \\
    {mohakg, lodewijk, namang, ashishg}@stanford.edu, kamesh@duke.cs.edu
}

\begin{document}

\maketitle

\begin{abstract}
Deliberative processes are often discussed as increasing or decreasing polarization. This approach misses a different, and arguably more diagnostic, dimension of opinion change: whether deliberation reshuffles who agrees with whom, or simply moves everyone in parallel while preserving the pre-deliberation rank ordering. We introduce \emph{opinion mixing}, measured by Kendall's rank correlation (\(\tau\)) between pre- and post-deliberation responses, as a complement to variance-based polarization metrics.
Across two large online deliberative polls spanning 32 countries (MCF-2022: \(n=6{,}342\); MCF-2023: \(n=1{,}529\)), deliberation increases opinion mixing relative to survey-only controls: treatment groups exhibit lower rank correlation on \(97\%\) and \(93\%\) of opinion questions, respectively. Polarization measures based on variance tell a more heterogeneous story: controls consistently converge, while treated groups sometimes converge and sometimes diverge depending on the issue.

To probe mechanisms, we link transcripts and surveys in a third event (SOF: \(n=617\), 116 groups) and use LLM-assisted coding of 6,232 discussion statements. Expressed support in discussion statements strongly predicts subsequent group-level opinion shifts; this correlation is amplified by justification quality in the statements but not by argument novelty. To our knowledge, we are the first to observe how different notions of argument quality have different associations with the outcome of deliberation. This suggests that opinion change after deliberation is related to selective uptake of well-reasoned arguments, producing complex patterns of opinion reorganization that standard polarization metrics may miss.
\end{abstract}

\section{Introduction}
Understanding how individuals or groups of people change their viewpoints upon deliberation is a central question in computational social science. 
Theories of deliberative democracy argue that well-structured discussion — where participants exchange reasons, justify their positions, and encounter diverse perspectives — produces better-informed and more reflective opinions. Fishkin's framework of Deliberative Polling institutionalized this vision by combining small-group deliberations and plenary expert panels with pre- and post-surveys to measure shifts in opinion~\citep{Fishkin2009WhenThePeopleSpeak}. A substantial literature has examined whether deliberation may alleviate polarization~\citep{fishkin2021deliberation, caluwaerts2023deliberation} or exacerbate it~\citep{Sunstein2002LawGroupPolarization, wojcieszak2011deliberation,gronlund2015does}, depending on group composition and deliberative context, typically by tracking changes in group means or variance of opinions.

Yet polarization and mean shifts, while important, capture only part of what deliberation might accomplish. Consider two convergence scenarios that are indistinguishable by just observing the opinion distribution: in the first, every participant shifts toward the group mean by the same proportion, preserving the rank order of who holds more or less extreme views; in the second, participants genuinely reconsider their positions, leading some who were initially skeptical to become supporters and vice versa. Both scenarios reflect fundamentally different processes. The first is consistent with social conformity or regression to the mean; the second suggests that deliberation facilitated genuine opinion revision—that exposure to arguments and evidence caused individuals to update their beliefs in ways that reshuffled their relative positions.

The same logic applies when opinions diverge. Consider two scenarios that produce identical increases in variance: in the first, every participant becomes more extreme in the direction they already leaned—supporters become stronger supporters, skeptics become more skeptical, amplifying pre-existing divisions. In the second, participants reshuffle: some former supporters develop doubts after hearing counterarguments, while some former skeptics become convinced by new evidence, and the individuals who end up at the extremes are not the same ones who started there. Both scenarios register as equivalent polarization under variance-based measures, yet they carry very different implications. The first reflects echo-chamber dynamics. The second, despite producing greater disagreement, reflects engagement: participants updated their views based on the discussion, and the resulting distribution, though more dispersed, emerged from genuine reconsideration.

This distinction matters for democratic theory. If deliberation merely compresses or stretches opinions while preserving relative ordering, it may reflect conformity or entrenchment rather than reasoned persuasion. However, if after deliberation, a rank reshuffling occurs, then it may suggest that participants are genuinely engaging with the substance of the discussion rather than simply drifting toward the median or reiterating prior positions. 

We propose to leverage this rank structure to characterize opinion change beyond polarization. We quantify \emph{opinion mixing} using Kendall's rank correlation ($\tau$) between pre- and post-deliberation surveys (Definition~\ref{def:kt}), which measures how much the relative ordering of individuals' opinions is preserved. A $\tau$ of $1$ indicates perfect preservation of ranks, while lower values indicate greater reshuffling. Critically, this metric is invariant to translation and scaling: a population that uniformly shifts in a certain direction, or whose variance shrinks or grows, will still show $\tau =1$ if no one changes position relative to anyone else. Rank correlation thus isolates the component of opinion change that opinion distribution-based measures cannot detect.

We complement opinion mixing with two additional metrics from prior literature that capture distinct phenomena. First, we examine \emph{variance change}, which measures whether the population opinion converged (reduced disagreement) or became more dispersed (increased disagreement) over the course of deliberation. Second, we assess \emph{softening of stances}, which we operationalize as mean reversion, or the negative correlation between individuals' initial opinions and their subsequent opinion changes; intuitively, this captures the extent to which those who started with more extreme views (higher or lower) tended to move toward the group mean. Together, these three metrics decompose opinion change into three parts: whether people moved relative to each other, whether the group as a whole became more or less dispersed, and whether those with extreme opinions softened their positions. 

These metrics characterize \textit{what} changes during deliberation; we also investigate \textit{why} such changes occur. Deliberative Polling events produce rich transcripts of small-group discussions, yet systematic analysis of these texts has remained limited by the prohibitive cost of manual coding. For the   \ifanonelse{\redacted}{\emph{Shaping Our Future}} (\textsc{SOF}) deliberation (Table~\ref{tab:events}), where we can link each room’s transcript to the corresponding pre- and post-survey responses, we apply an LLM annotator rubric adapted from prior work \citep{gelauff2024estimating} to annotate each statement for (i) stance (expressed support or opposition toward the agenda), (ii) justification quality (examples and anecdotes to support the statement), and (iii) novelty (new information relative to the prior statements made in the deliberation group). We aggregate these annotations into group-level features and link them to pre--post opinion shifts at the deliberation group level, controlling for baseline opinions and topic fixed effects. This approach not only deepens our understanding of the mechanics of deliberation but also provides actionable insights for designing more effective deliberative polling events that foster better engagement and informed decision-making.

\paragraph{Three Deliberation Events.}
Our datasets come from three deliberative polling events run on the \ifanonelse{\redacted}{Stanford} Online Deliberation Platform: \ifanonelse{\redacted}{Meta} Community Forum 2022 (\textsc{MCF-2022}), \ifanonelse{\redacted}{Meta} Community Forum 2023 (\textsc{MCF-2023}), and \ifanonelse{\redacted}{\emph{Shaping Our Future}} (\textsc{SOF}). Table~\ref{tab:events} summarizes their scale (participants, countries/languages, and number of small groups), format (sessions per participant), and survey design (number of opinion questions and the prevalence of ``no opinion'' responses pre/post).  Both MCF datasets include a control group that completed the same pre--post surveys, enabling treatment--control comparisons. This group did not receive the briefing material or participate in any way other than responding to both surveys. \textsc{SOF} did not include a control group, but it provides connected transcript data that we use to study the mechanisms of opinion change.

Across events, opinion questions were answered on a 0--10 integer scale with an explicit ``no opinion'' option. The pre-survey was administered before any briefing materials, whereas the post-survey was given after the final plenary session. We excluded demographic and identity questions to maintain an identity-agnostic analysis. Further details of these datasets and the deliberation platform are in Section~\ref{sec:data}.

\subsection{Overview of Results}

To isolate the effects of deliberation from simple re-measurement (the so-called panel conditioning effect \citep{WarrenHalpernManners2012SMR,SturgisAllumBruntonSmith2009PanelConditioning,PEW2021PanelConditioning}) or time effects, we compare the treatment set to the control set’s repeated polls. The effect observed in the control group corresponds to the case where participants reconsider their views independently after the initial opinion survey. We find that panel conditioning generally reduces the variance of opinions.

In contrast, \emph{deliberation} leads to more complex outcomes in our dataset. Kendall’s $\tau$ \citep{abdi2007kendall} declines relative to the control set for most questions, implying greater reordering of opinions. Variance sometimes increases and sometimes decreases, depending on the issue, suggesting that after deliberation, both increased consensus and sharpened disagreement can occur. This variability complicates narratives that equate deliberation purely with convergence or polarization. Participants also exhibit marginally greater softening of stances, i.e., higher mean reversion than the control set, indicating that exposure to a balanced set of arguments prompts moderation of extremes.

\paragraph{Mechanisms of Opinion Change.}
Three results stand out in the SOF dataset. First, greater expressed support within a group is associated with larger movement toward that stance on the corresponding agenda; the association extends, albeit slightly weakly, to participants who do not speak on that topic, suggesting influence through exposure. Second, justification quality enhances this relationship: groups with higher average reason-giving show stronger alignment between expressed support and opinion change. Third, we do not observe a comparable enhancing role for novelty, suggesting that whether participants justify positions matters more than whether they introduce new information per se. These findings suggest a design implication: deliberation practices should prioritize structured reason-giving and high-quality justifications over mere information sharing. While our analyses are observational and do not identify causal effects, the patterns are robust and hold after adjusting for baseline opinions and topic-level heterogeneity.


\section{Related Work}
\label{sec:related}

\textbf{Deliberation and Deliberative Polling.}
Deliberative democracy grounds legitimacy in reason-giving among free and equal citizens \citep{Habermas1984,Cohen1989,GutmannThompson1996}. 
Deliberative polling operationalizes this ideal through briefing materials, moderated small-group discussions, expert panels, and pre- and post-measurement relative to a control \citep{Fishkin2009WhenThePeopleSpeak,fishkin2019deliberative}. Prior studies report that deliberation can increase knowledge and shift opinions, with variance sometimes rising (disagreement) and sometimes falling (consensus) \citep{FishkinLuskin2005DelibPolling,luskin2004considered,luskin2022deliberative}.
\citet{WarrenPearse2008} and \citet{FarrellEtAl2019} provide design principles for citizens' assemblies and other deliberative mini-publics that inform our platform architecture. The migration of deliberative practices to digital platforms raises questions about quality and outcomes. \citet{gelauff2023achieving} show that online deliberations can achieve parity with in-person deliberation in terms of participant satisfaction and equity of speaking turns.

\textbf{Deliberation and Polarization.} A substantial body of literature examines the relationship between polarization and deliberation. \citet{Sunstein2002LawGroupPolarization} finds that deliberation between like-minded people causes their opinions to become even more extreme. \citet{caluwaerts2023deliberation} does a multidisciplinary meta-analysis of eighty research papers studying this connection, and found a roughly even split between those reporting an increase in polarization and those reporting a decrease. 
Perhaps ironically, there is no consensus on the best mathematical definition of polarization in this context. \citet{bramson2017understanding} gives nine definitions which may be applicable to different scenarios. These include the spread, variance, coverage, and measures of bimodality and multimodality of the opinion distribution. There are some definitions in the literature which require an assigned group identity (e.g. political left or right), and correspond to tracking how the opinions of each group moves. These measures lose the information of relative reordering at the participant level that is captured by Kendall's $\tau.$ To the best of our knowledge, ours is the first work to study Kendall's $\tau$ in the context of deliberation-induced opinion change. 

\textbf{Mechanisms of Opinion Change.} To study how specific features of the conversation affect opinion change, we must be able to reliably annotate the transcript for those features. Work on discourse quality emphasizes justification and respect \citep{SteenbergenEtAl2003} and highlights the roles of narrative and emotion \citep{BachtigerEtAl2010}.
A growing literature evaluates text with LLMs. 
\citet{ZhengEtAl2023MTBench} show strong agreement between GPT-4 and human preferences on open-ended dialogue tasks. \citet{gelauff2024estimating} demonstrated that LLMs can match or exceed human reliability at evaluating the quality of deliberation statements.

\citet{Neblo2005ChangeBetter} argues that the case for deliberative reforms hinges on whether opinion change occurs via mechanisms congruent with normative theories (e.g., reason-giving, reciprocity). From five small-group deliberations, they found that deliberative influence was mediated by `respect' social networks, and not friendship and familiarity.
In more recent work, \citet{nakazawa2024effect} found that sharing \textit{personal experiences} has a significant impact on opinion change in a study conducted with sixty-five participants who deliberated in eight groups. Similarly, in a study of thirteen small-group deliberations, \citet{gerber2014deliberative} found that statements backed by reason resulted in more opinion change.  

Our paper extends this line of work in the following ways. (1) By using LLMs as annotators, we study this question at a much larger scale, thereby obtaining more precise estimates. (2) Unlike prior work, we perform this analysis on online deliberation, which has notably different conversation dynamics (including reduced nonverbal cues) \citep{BaekEtAl2012}. (3) Finally, unlike prior work, we consider the two dimensions of deliberative quality, novelty and justification, and highlight the difference in how these are associated with changes in the opinion. 

\textbf{Mathematical Models of Deliberation and Opinion Change.}
Classic linear-averaging models predict convergence to consensus \citep{DeGroot1974Consensus}. \citet{FriedkinJohnsen2011SIN} extend this to allow persistent disagreement with ``stubbornness.'' 
Biased assimilation on homophilous networks yields polarization when individuals overweight congenial evidence \citep{DandekarGoelLee2013PNAS}. Our work challenges these models of opinion change, many of which fail to explain opinion mixing and overlook the relationship between discourse quality and opinion change.

Several game-theoretic models study deliberation before voting and analyze how communication affects information aggregation and outcomes \citep{AustenSmithFeddersen2006APSR, GerardiYariv2007JET,landa2009game}. 
\citet{goel2025metric} shows that deliberation before voting can improve the social utilities of election outcomes. \citet{meirowitz2007defense} studies a model of strategic communication and voting and finds that small groups lead to truthful information sharing and bigger groups may not. These works underline the positive impact of deliberation in collective decision-making.

\section{Data and Deliberative Polling Setup}
\label{sec:data}

Our data come from three major deliberative polling events conducted on the \emph{\ifanonelse{\redacted}{Stanford} Online Deliberation Platform}. The platform enables automated moderation in parallel small-group deliberations over structured agendas, reproducing the essential features of in-person deliberative polls in an online environment \citep{gelauff2023achieving}. Each deliberation event consists of several nested organizational units: \emph{rooms}, \emph{roomgroups}, \emph{sessions}, and \emph{deliberation events}.  

A \emph{room} is a small group of typically 5--12 participants who deliberate on a shared agenda. Several rooms running in parallel form a \emph{roomgroup}, all using identical agendas and platform configurations. A \emph{session} can consist of one or multiple roomgroups that may meet at different times or with different configurations, to accommodate, for example, participant availability or language preferences. Finally, a \emph{deliberation event} consists of one or more sessions under a common set of topics and procedures, typically alternating with plenary expert panels.

Participants are recruited through professional polling firms to ensure representativeness of the target population, and are given reasonable monetary compensation for their time. Assignment to treatment or control is at random. Before small-group deliberations, they receive balanced briefing materials that summarize the arguments for and against each proposal on the agenda, which typically contains 3--6 items. Participants are randomly assigned to small-group deliberation rooms at the start of the session, subject to minimum and target room sizes.

The automated moderator manages speaking turns using a visible queue: each speaker may talk for up to 45 seconds before the next queued participant’s turn begins. Groups can vote to move to the next agenda item once discussion is deemed sufficient. After all agenda items are covered, participants draft, edit, and rank questions for an expert panel.
We focus on three deliberation events. Table~\ref{tab:events} summarizes the three events. We describe each briefly below.

\textbf{MCF-2022.} Conducted in 2022 in collaboration with \ifanonelse{\redacted}{Meta Inc.}, this event spanned 32 countries and 19 languages. Discussions centered on online harm, safety, and governance in the emerging \ifanonelse{\redacted}{Metaverse}. Each participant joined up to four small-group deliberations. The 71 opinion questions used a 0–10 integer scale with a ``no opinion'' option. The agenda items are in Appendix~\ref{app:agendas}.

\textbf{MCF-2023.} Conducted in 2023, also in collaboration with \ifanonelse{\redacted}{Meta Inc.}, this event focused on generative AI policies. Participants again joined up to four small-group sessions. The 44 opinion questions followed the same format.

\textbf{SOF.} Conducted in 2021 with 38 U.S. post-secondary institutions, this event involved two deliberation sessions per participant, one on political reforms (60 rooms) and one on climate and economic reforms (56 rooms). Eight survey questions corresponded directly to the agenda items (see Appendix~\ref{app:agendas}). Since our goal is to link deliberation features to opinion change, we restrict our attention to these agenda-matched questions and omit those about related issues.

Since SOF lacked a control group, we cannot analyze opinion mixing relative to a no-deliberation counterfactual for that event. However, we use its transcripts to study the mechanisms of opinion change. For MCF-2022 and MCF-2023, we lack a clean correspondence between transcripts and survey questions, so we use those datasets only for studying aggregate opinion change. Technical reports for each event provide question-level results and demographic breakdowns \ifanonelse{\redacted}{\citep{fishkin2023meta, chang2024meta, StanfordDDL2021ShapingOurFutureReport}}. The data for this study were collected with the explicit consent of the survey respondents and deliberation participants.  

\begin{table*}[t]
\centering
\small
\begin{tabular}{@{}p{3.3cm} ccc ccc c@{}}
\toprule
& \multicolumn{3}{c}{\textbf{MCF-2022}} & \multicolumn{3}{c}{\textbf{MCF-2023}} & \textbf{SOF} \\
\cmidrule(lr){2-4} \cmidrule(lr){5-7} \cmidrule(lr){8-8}
& Treatment & Control & & Treatment & Control & & Treatment \\
\midrule
Participants & 6,342 & 6,101 & & 1,529 & 1,094 & & 617 \\
Countries & \multicolumn{2}{c}{32} & & \multicolumn{2}{c}{4} & & 1 \\
Languages & \multicolumn{2}{c}{19} & & \multicolumn{2}{c}{4} & & 1 \\
Small groups & \multicolumn{2}{c}{2,069} & & \multicolumn{2}{c}{792} & & 116 \\
Sessions/participant & \multicolumn{2}{c}{$\leq 4$} & & \multicolumn{2}{c}{$\leq 4$} & & 2 \\
\addlinespace[4pt]
Opinion questions & \multicolumn{2}{c}{71} & & \multicolumn{2}{c}{44} & & 8 \\
\addlinespace[4pt]
\multicolumn{8}{@{}l}{\textit{``No opinion'' fraction (\%)}} \\
\quad Pre-survey & 7.3 & 9.4 & & 9.2 & 8.3 & & 14.0 \\
\quad Post-survey & 3.3 & 7.6 & & 5.9 & 8.2 & & 3.2 \\
\addlinespace[4pt]
Key topics & \multicolumn{3}{c}{\parbox[t]{4.5cm}{\centering Online safety, \ifanonelse{\redacted}{Metaverse} governance}} 
           & \multicolumn{3}{c}{\parbox[t]{4.5cm}{\centering Generative AI and its engagement with users}} 
           &  \parbox[t]{4cm}{\centering Political reforms, climate, economic reform} \\
\bottomrule
\end{tabular}
\caption{Overview of deliberation events. The respondent count in each group corresponds to those who submitted both surveys. Observe that response rate increases after deliberation, indicating that deliberation is effective at helping people form opinions.}
\label{tab:events}
\end{table*}

\section{Methodology}
\label{sec:method}

\subsection{Notation and Preliminaries}
Let \(n\) denote the number of respondents in the relevant set (treatment or control), and $p$ the number of questions in the opinion survey.  For respondent \(i\) and survey item \(j\) let \(y_{i,j}^{\mathrm{pre}}\in\{0,1,\dots,10\}\) and \(y_{i,j}^{\mathrm{post}}\in\{0,1,\dots,10\}\) denote their integer pre- and post-deliberation opinions. A `no opinion' response is modeled as a missing value. For each question, we only consider those responses that have an integer response in both surveys when calculating our metrics. We assume that this does not introduce a selection bias in our results. Relaxing this assumption can be the subject of future work. 
While opinion \textit{formation} (going from ``no opinion'' to a numerical response) is interesting in itself \citep{SturgisAllumBruntonSmith2009PanelConditioning}, we cannot include these respondents in the metrics that we study in this paper. 

Let $n_j$ denote the number of respondents who give an integer response to item $j$ in both surveys. 
Define the opinion change as
\(
\Delta_{i,j} \;=\; y_{i,j}^{\mathrm{post}} - y_{i,j}^{\mathrm{pre}}.
\) Let \(\bar y_j^{\mathrm{pre}}\), \(\bar y_j^{\mathrm{post}}\) and \(\overline{\Delta}_j\) be the sample means of \(y_{\cdot,j}^{\mathrm{pre}}\), \(y_{\cdot,j}^{\mathrm{post}}\) and \(\Delta_{\cdot,j}\), respectively.

We measure disagreement on item \(j\) by the sample variance across respondents,
\[
\operatorname{Var}_j^{\mathrm{pre}} =   \frac{\sum\limits_{i=1}^{n_j} \bigl(y_{i,j}^{\mathrm{pre}} - \bar y_j^{\mathrm{pre}}\bigr)^2}{n_j-1},
\operatorname{Var}_j^{\mathrm{post}} = \frac{\sum\limits_{i=1}^{n_j} \bigl(y_{i,j}^{\mathrm{post}} - \bar y_j^{\mathrm{post}}\bigr)^2}{n_j-1}.
\]
\begin{definition} [Change in Disagreement]
    The relative increase in disagreement upon deliberation is the percentage change in variance:
\[
\Delta\mathrm{Var}_j \;=\; \frac{\operatorname{Var}_j^{\mathrm{post}} - \operatorname{Var}_j^{\mathrm{pre}}}
{\operatorname{Var}_j^{\mathrm{pre}}}\times 100\%.
\]
\end{definition} 

 We define \emph{stance softening} as the negative Pearson correlation between the pre-deliberation opinion and the change in opinion.
 
\begin{definition} [Stance Softening] \label{def:mr}
The mean-reversion (stance softening) for item \(j\) is the negative correlation of the pre-opinions and the opinion change, that is:
\[
\mathrm{MR}_j \;=\; -\frac{\displaystyle\sum_{i=1}^{n_j}\bigl(y_{i,j}^{\mathrm{pre}}-\bar y_j^{\mathrm{pre}}\bigr)\bigl(\Delta_{i,j}-\overline{\Delta}_j\bigr)}
{\sqrt{\displaystyle\sum_{i=1}^{n_j}\bigl(y_{i,j}^{\mathrm{pre}}-\bar y_j^{\mathrm{pre}}\bigr)^2}\,
\sqrt{\displaystyle\sum_{i=1}^{n_j}\bigl(\Delta_{i,j}-\overline{\Delta}_j\bigr)^2}},
\]
\end{definition}
Positive values of \(\mathrm{MR}_j\) indicate that higher pre-survey opinions tend to decrease and lower pre-survey opinions tend to increase, i.e., movement toward the center of the opinion distribution. Some positive mean-reversion is expected due to the inherent randomness in survey data. In the extreme case, if both pre- and post-survey responses were independent uniform random variables in $\{0,1,\ldots, 10\},$ we would have a mean reversion of $\frac{1}{\sqrt{2}}.$ This makes comparison with a no-deliberation control set crucial for interpretation.

We quantify population-level \emph{opinion mixing} by the change in rank-order agreement between the pre- and post-surveys for each item using Kendall's rank correlation coefficient. We first do a pre-processing step to ensure consistent tie-breaking.
To produce stable rank statistics, we add a small uniform random perturbation to each participant's response vector before calculating Kendall's  $\tau$.
\[
\tilde y_{i,j}^{\mathrm{pre}} \;=\; y_{i,j}^{\mathrm{pre}} + \varepsilon_i,\qquad
\tilde y_{i,j}^{\mathrm{post}} \;=\; y_{i,j}^{\mathrm{post}} + \varepsilon_i,
\]
where \(\varepsilon_i \sim \mathrm{Uniform}(0,\,0.1)\) and the same draw \(\varepsilon_i\) is used for participant \(i\) in both pre and post. This preserves ordinal comparisons while breaking ties \emph{consistently} between pre and post. The number $0.1$ is arbitrary, and we are theoretically guaranteed to get identical results for any value in the interval $(0,1).$ This technique ensures that for two participants with identical scores in both surveys, they are not counted as a rank inversion pair. 

\begin{definition}[Kendall's Rank Correlation ($\tau$) \citep{abdi2007kendall}]\label{def:kt}
    For item \(j\) consider the two vectors of (tie-broken) opinions across participants \(\tilde{\mathbf y}^{\mathrm{pre}}_{\,\cdot,j}\) and \(\tilde{\mathbf y}^{\mathrm{post}}_{\,\cdot,j}\). Kendall's \(\tau\) is:
    \[
\tau_j \;=\; \frac{1}{\binom{{n_j}}{2}} \sum_{1\leq a < b \leq {n_j}} \operatorname{sign}\bigl(\tilde y_{a,j}^{\mathrm{pre}} - \tilde y_{b,j}^{\mathrm{pre}}\bigr)\,\operatorname{sign}\bigl(\tilde y_{a,j}^{\mathrm{post}} - \tilde y_{b,j}^{\mathrm{post}}\bigr).
\]
\end{definition}
Smaller values of \(\tau_j\) correspond to greater reordering (more mixing) between pre and post ranks. Same as mean-reversion, some amount of rank-inversions are expected due to randomness in the data. Therefore comparison with a control set is crucial to get meaningful insights.

\subsection{Coding of Statements in Deliberation Transcripts}

To connect discussion features with opinion change, we code each statement in the SOF transcripts along deliberative quality (novelty and justification) and stance. We follow the methodology of \citet{gelauff2024estimating}, who used large language models (LLMs) to code deliberative quality. The novelty score assesses whether: ``This statement introduces novel ideas, perspectives, or solutions.'' The justification score assesses whether: ``This statement includes examples or anecdotes to support the speaker’s point.'' We use the same model (GPT-4 from OpenAI), rubric, and prompt without any few-shot examples. Both metrics are scored on a five-point Likert scale (1–5), where higher values denote greater justification quality or novelty.

In addition to these metrics, we introduce \emph{expressed support}, which captures whether a statement supports, opposes, or neutrally discusses the agenda item under deliberation. Expressed support is coded as an integer in $\{0,1,2\}$, corresponding respectively to opposition, neutrality, and support. 

To code the expressed support, we augment the prompt with six agenda-specific few-shot examples that illustrate the full range of possible responses. For each agenda item, we hand-picked these examples from the corpus of the transcripts to be instructive and representative. Each example statement is followed by a model justification to demonstrate the reasoning process expected from the LLM. The final prompt requests both a rating and a short textual justification. The addition of few-shot examples helps ensure semantic consistency in how the model interprets stance intensity across agenda items. We use the newer GPT-4o model from OpenAI for annotating the stance.

\subsection{Human Validation of Stance Annotations}
\label{sec:stance_validation}

To validate our LLM-based stance labels, we conducted a human annotation study with four graduate-student raters who independently labeled 40 randomly sampled statements on the same three-point scale (oppose/neutral/support). Human inter-rater reliability was moderate: Krippendorff's $\alpha$ (ordinal) $= 0.558$ and Fleiss' $\kappa = 0.397$, with complete four-rater agreement on only 37.5\% of statements. This confirms that stance attribution at the statement level is inherently subjective. To evaluate LLM performance, we used a leave-one-rater-out design comparing each rater's deviation from the remaining three-rater consensus against the LLM's deviation from the same target. The LLM matched or exceeded the held-out human on 94\% of comparisons (150 of 160 statement--rater pairs were ties or LLM wins; binomial $p < 0.001$ excluding ties). These results indicate that our LLM annotations track human consensus at least as reliably as individual raters, supporting their use at scale while reinforcing the need for aggregation across statements to obtain stable group-level measures. Full details of the validation experiment are in Appendix~\ref{app:stance_validation}.

\subsection{Linking Discussion Features to Opinion Change}

To estimate how discussion features relate to opinion change, we conduct an ordinary least squares (OLS) regression at the level of \emph{room--agenda item pairs}. Each data point corresponds to one agenda item discussed in one deliberation room. The first three agenda items in SOF were discussed in 60 rooms and the next five items in 56 rooms, resulting in a total of 460 observations across 116 room configurations.
For each room--agenda item pair, we compute:
\begin{itemize}[leftmargin = 12pt]
    \item the mean pre- and post-deliberation opinions across all participants in the room for that agenda item, and define the dependent variable as the change;
    \item the mean \emph{novelty}, \emph{justification,} and \emph{expressed support} scores across all statements made in that room on that agenda item.
\end{itemize}

We mean-center all continuous predictors prior to constructing interaction terms to reduce multicollinearity. We use heteroscedasticity-consistent (HC3) robust standard errors to account for potential heterogeneity in the variance of the residuals. We also compute variance inflation factor (VIF) scores for the predictors to assess multicollinearity, since high collinearity can inflate standard errors and make coefficient estimates unstable or difficult to interpret. All VIF scores were less than 2, indicating the absence of multicollinearity.
The regression model is specified as follows:
\[
\begin{aligned}
&\Delta o_{ri} =\; \beta_0 
+ \beta_1 S_{ri} 
+ \beta_2 N_{ri}
+ \beta_3 J_{ri}
+ \beta_4 (S_{ri} \times N_{ri}) \\
&+ \beta_5 (S_{ri} \times J_{ri})
+ \beta_6 O_{ri} 
+ \beta_7 L_{ri}
+ \sum_{k=2}^{K} \gamma_k A_{k,ri}
+ \epsilon_{ri}.
\end{aligned}
\]
where $\Delta o_{ri}$ denotes the average opinion change for room $r$ on agenda item $i$, $S_{ri}$ the mean expressed support, $N_{ri}$ the mean novelty, $J_{ri}$ the mean justification, $O_{ri}$ the average opinion pre-deliberation, $L_{ri}$ the log number of statements, and $A_{k,ri}$ the  dummy variables for each agenda item $k$ (with one omitted as the reference). The variable $A_{k,ri}$ is one for a sample corresponding to agenda k and zero for all other samples. Including these controls is essential to account for baseline differences in opinions (e.g., regression to the mean from \(O_{ri}\)), discussion length, and agenda-specific effects, ensuring a more accurate estimation of the relationships between discussion features and opinion change. 

We add the log of the number of statements, rather than the number of statements, because the effect of the length of the discussion on opinion change saturates after a certain point, and the distribution of the number of statements on an agenda item exhibits a large spread. 

Observe that we do not add room fixed effects in the regression. This is because each room contributes only 3 or 5 data points (one per agenda item), and including this effect could reduce the power of the result. This is reasonable since the automated facilitation and random room assignment are assured on the deliberation platform. However, we acknowledge that unmeasured room-level factors, such as the particular mix of personalities or idiosyncratic group dynamics, could still introduce bias that our controls do not capture, and this could be a subject for further research.

This is a \emph{moderation analysis} \citep{aiken1991multiple}, and examines whether the relationship between an independent variable and an outcome depends on the value of a third variable, the moderator. In our setting, we test whether discourse quality features such as justification and novelty \emph{moderate} the association between expressed support and opinion change. Positive interaction coefficients indicate that higher quality amplifies the relation between expressed support and opinion change, whereas negative coefficients suggest attenuation.

Since deliberation and opinion change are not experimentally separated, the observed relationships cannot be interpreted causally. Opinion change could reflect prior reflection or reactions to briefing materials, which in turn influenced the stance participants expressed in the discussion. To test robustness, we repeat the analysis, restricting the outcome variable to the opinion change of participants who did not contribute any statement on that agenda item. An average of 28.4\% of participants do not make any statement in a given room-agenda pair. The distribution of this fraction is in Figure~\ref{fig:observer_dist} in Appendix~\ref{app:fig}.

\section{Results}
\label{sec:results}
\subsection{Characterizing Opinion Change }
We analyze opinion change across 71 policy items in MCF-2022 and 44 policy items in MCF-2023. The control group isolates the effect of panel conditioning, allowing us to attribute differences to the deliberation event.

For each metric, we compute item-level statistics separately for treatment and control groups, then conduct inference across items. For each item, we only consider responses that provide an integer score in both the pre- and post-surveys. We report medians rather than means as our primary measure of central tendency, given the potential for outliers in item-level effects. Confidence intervals for data set statistics (95\%) are obtained via bootstrap resampling over items (10,000 iterations). We assess statistical significance using paired $t$-tests across items, with $p$-values reported in Table~\ref{tab:opinion_dynamics}. Wilcoxon signed-rank tests yield consistent conclusions in all cases, confirming that results are not sensitive to distributional assumptions. We report effect sizes as Cohen's $d$, computed as the mean paired difference divided by the standard deviation of paired differences. For variance change, where our interest lies in the \emph{dispersion} of effects rather than central tendency alone, we additionally compare the standard deviation of variance changes across items between treatment and control using the Brown-Forsythe test. We also report the proportion of items showing effects in the predicted direction (e.g., $\tau_{\text{treatment}} < \tau_{\text{control}}$), which provides an intuitive measure of effect consistency that is robust to outliers.

\begin{figure*}[t]
\centering
\begin{subfigure}[b]{0.32\textwidth}
    \centering
    \includegraphics[width=\textwidth]{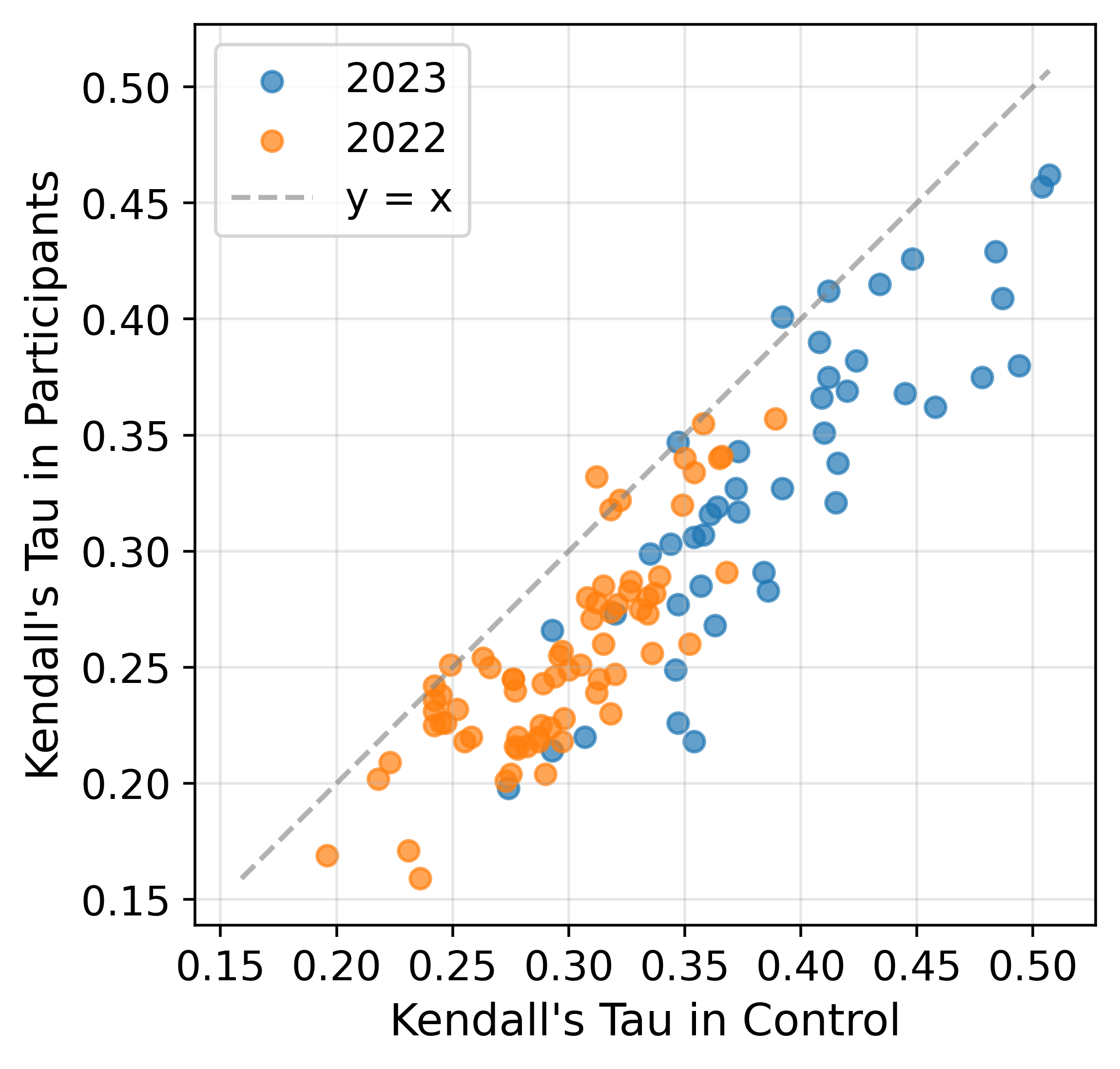}
    \caption{Kendall's $\tau$}
    \label{fig:kt_meta}
\end{subfigure}
\hfill
\begin{subfigure}[b]{0.32\textwidth}
    \centering
    \includegraphics[width=\textwidth]{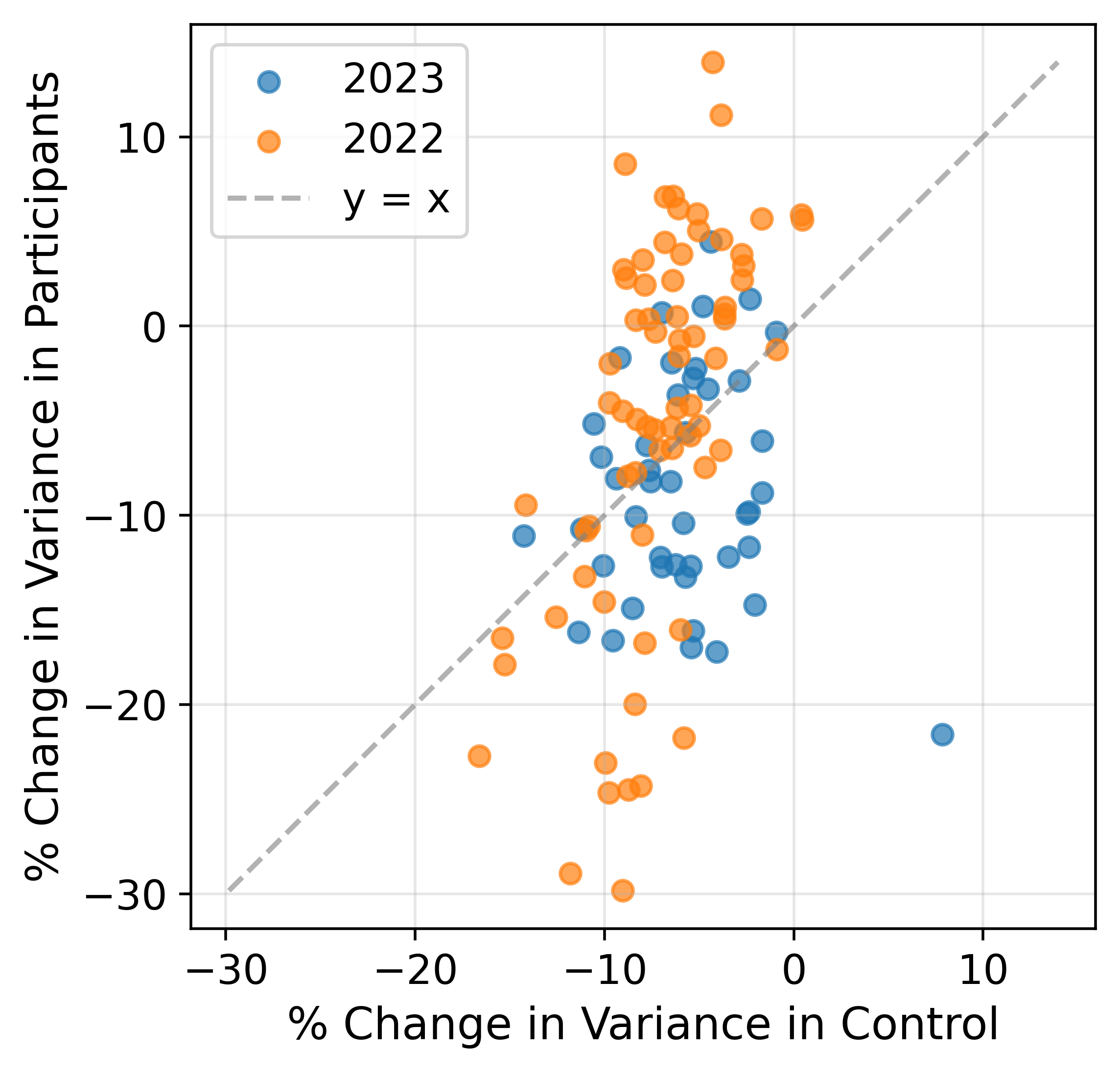}
    \caption{Percentage change in variance}
    \label{fig:var_meta}
\end{subfigure}
\hfill
\begin{subfigure}[b]{0.32\textwidth}
    \centering
    \includegraphics[width=\textwidth]{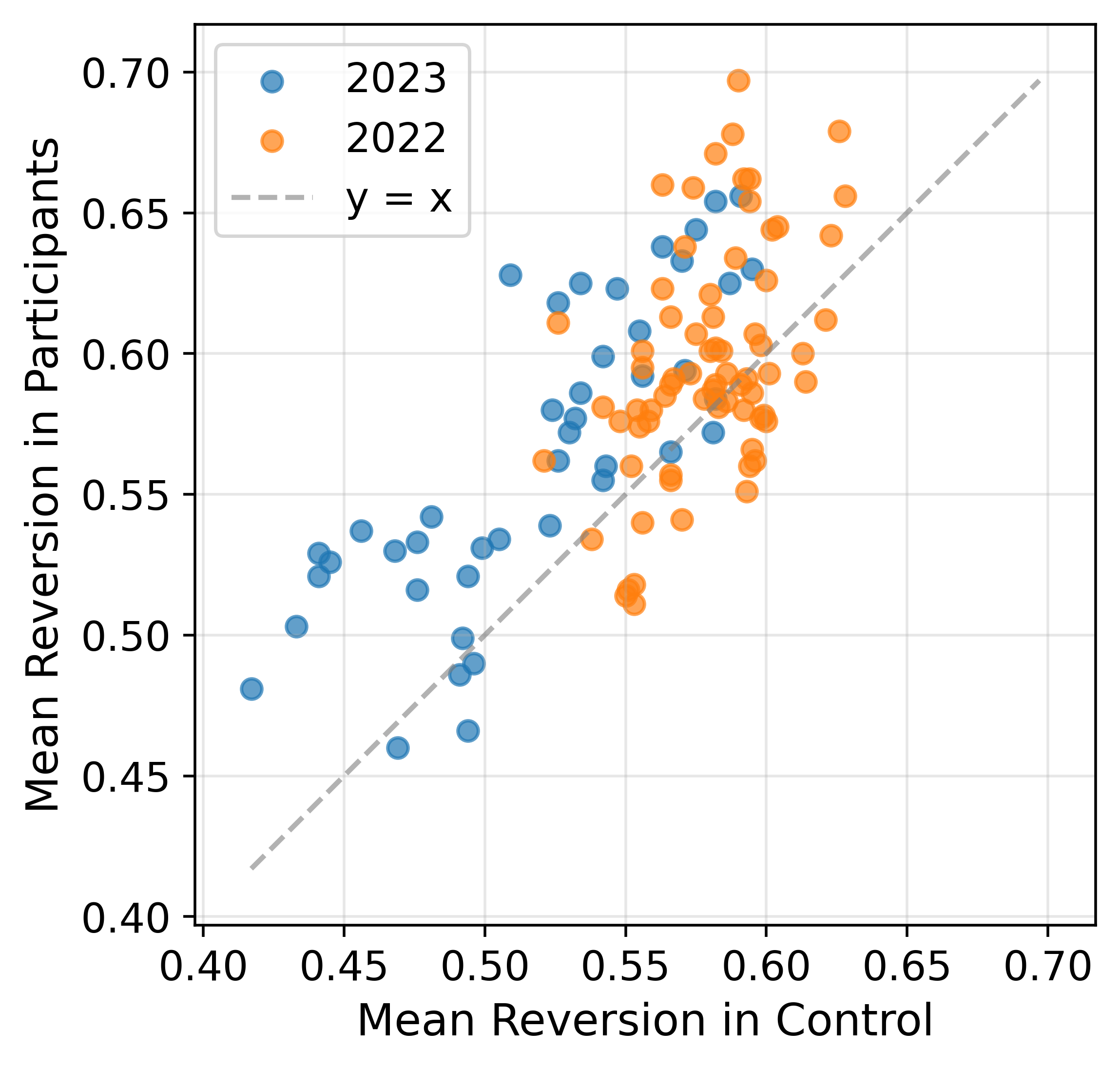}
    \caption{Mean reversion}
    \label{fig:meanrev_meta}
\end{subfigure}

\caption{Treatment vs.\ control comparisons. Each point represents one item. 
(a)~Kendall's $\tau$ between pre- and post-deliberation opinions; nearly all points lie below the 
diagonal, indicating greater opinion mixing in the treatment group. 
(b)~Percentage change in opinion variance; control items are in a narrow band of mild decreases, 
while treatment items show heterogeneous effects spanning both convergence and divergence. 
(c)~Mean reversion; more points lie above the diagonal, indicating modestly greater moderation of 
extreme opinions in the treatment group.}
\label{fig:meta_analysis}
\end{figure*}

\begin{table*}[t]
\centering
\small
\begin{tabular}{@{}llcccc@{}}
\toprule
\textbf{Event} & \textbf{Metric} & \textbf{Treatment} & \textbf{Control} & \textbf{Median Diff.} & \textbf{$p$} \\
\midrule
\multirow{4}{*}{MCF-2022} 
& Kendall $\tau$ & 0.248 \, [0.237, 0.258] & 0.296 \, [0.286, 0.312] & $-0.043$ & $<.001$ \\
& Variance change (\%) & $-4.0$ \, [$-5.5$, 0.3] & $-6.8$ \, [$-8.0$, $-6.1$] & $+4.0$ & .037 \\
& \quad SD across items (\%) & 10.1 \, [8.4, 11.5] & 3.4 \, [2.7, 4.0] & $+6.8$ & $<.001$ \\
& Mean reversion & 0.591 \, [0.584, 0.601] & 0.582 \, [0.573, 0.589] & $+0.019$ & $<.001$ \\
\addlinespace
\multirow{4}{*}{MCF-2023} 
& Kendall $\tau$ & 0.327 \, [0.306, 0.365] & 0.385 \, [0.361, 0.411] & $-0.053$ & $<.001$ \\
& Variance change (\%) & $-9.3$ \, [$-11.5$, $-6.5$] & $-5.8$ \, [$-7.0$, $-5.2$] & $-1.7$ & .014 \\
& \quad SD across items (\%) & 5.9 \, [4.8, 6.8] & 3.7 \, [2.5, 4.8] & $+2.2$ & $<.001$ \\
& Mean reversion & 0.564 \, [0.536, 0.586] & 0.526 \, [0.496, 0.542] & $+0.048$ & $<.001$ \\
\bottomrule
\end{tabular}

\raggedright
\caption{Summary of opinion dynamics metrics across 71 items (MCF-2022) and 44 items (MCF-2023). For Kendall $\tau$ and mean reversion, we report medians with bootstrapped 95\% confidence intervals. For variance change, we report both the median change and the standard deviation across items to capture heterogeneity. Statistical tests compare treatment to control: paired $t$-tests for central tendency, Brown-Forsythe for dispersion.}
\label{tab:opinion_dynamics}
\end{table*}

\subsubsection{Opinion Mixing: Widespread Reordering After Deliberation}

Figure~\ref{fig:kt_meta} displays the per-item Kendall's \(\tau\) between pre- and post-survey responses. Across 71 opinion items in MCF-2022,
the treatment group showed lower rank correlation between pre- and post-deliberation opinions
(median \(\tau = 0.248\), 95\% CI \([0.237, 0.258]\)) than the control group (median \(\tau = 0.296\),
95\% CI \([0.286, 0.312]\)). The median paired difference was \(-0.043\) (95\% CI \([-0.049, -0.034]\)),
with treatment \(\tau\) lower than control \(\tau\) on 69 of 71 items (97\%). A paired \(t\)-test across items confirmed this difference
(\(t = -13.71\), \(p < 0.001\), Cohen's \(d = -1.64\)).

The MCF-2023 results show the same pattern. Across 44 opinion items, the treatment group showed lower
rank correlation between pre- and post-deliberation opinions (median \(\tau = 0.327\), 95\% CI
\([0.306, 0.365]\)) than the control group (median \(\tau = 0.385\), 95\% CI \([0.361, 0.411]\)).
The median paired difference was \(-0.053\) (95\% CI \([-0.074, -0.045]\)), with treatment \(\tau\) lower
than control \(\tau\) on 41 of 44 items (93\%) (two ties; one item with higher \(\tau\) in treatment).
A paired \(t\)-test across items confirmed this difference (\(t = -11.97\), \(p < 0.001\), Cohen's \(d = -1.83\)).

\subsubsection{Variance Change: Heterogeneity Versus Uniformity}
Deliberation did not uniformly reduce or increase disagreement; instead, it amplified the variance change in both directions. 
Figure~\ref{fig:var_meta} plots the percentage change in within-group variance for each item. 
In MCF-2022, while the median variance change was similar between groups (treatment: \(-4.0\%\), 
95\% CI \([-5.5\%, 0.3\%]\); control: \(-6.8\%\), 95\% CI \([-8.0\%, -6.1\%]\)), the treatment group showed 
substantially more heterogeneous effects across items. The standard deviation of variance changes was 
threefold larger in treatment (SD \(= 10.1\%\), 95\% CI \([8.4\%, 11.5\%]\)) than in control 
(SD \(= 3.4\%\), 95\% CI \([2.7\%, 4.0\%]\)), a difference that was highly significant 
(Brown-Forsythe \(F = 48.96\), \(p < 0.001\)). In treatment, 28 items (39\%) showed increased variance 
while 43 (61\%) showed decreased variance; in control, only 2 items (3\%) showed any variance increase, both with less than 1\% increase.

The MCF-2023 results exhibit a similar pattern. The median variance change was \(-9.3\%\) (95\% CI \([-11.5\%, -6.5\%]\)) in treatment and \(-5.8\%\) (95\% CI \([-7.0\%, -5.2\%]\)) in control. Treatment again showed greater heterogeneity: the standard deviation of variance changes was 1.6-fold larger in treatment (SD \(= 5.9\%\), 95\% CI \([4.8\%, 6.8\%]\)) than in control (SD \(= 3.7\%\), 95\% CI \([2.5\%, 4.8\%]\)), a significant difference (Brown-Forsythe \(F = 11.83\), \(p < 0.001\)). In treatment, 4 items (9\%) showed increased variance while 40 (91\%) showed decreased variance; in control, only 1 item (2\%) showed any variance increase. Among treatment items, 20 (45\%) showed strong convergence (variance decrease exceeding 10\%), compared to only 6 (14\%) in control.

This heterogeneity complicates simplistic narratives equating deliberation with either consensus or polarization. Instead, the data suggest that structured discussion can both \emph{integrate} views on some issues and \emph{differentiate} them on others, depending on the arguments raised, the distribution of prior beliefs, and the normative contestability of the issue.

\subsubsection{Mean Reversion: Modest Evidence of Opinion Softening}
Participants with initially extreme opinions reversed to the mean more after deliberation than in the control condition, though the effect is modest. 
Figure~\ref{fig:meanrev_meta} displays the mean reversion statistic (negative correlation between pre-survey opinion and opinion change) for each item. 
In MCF-2022, deliberating respondents showed slightly stronger mean reversion (median \(= 0.591\), 95\% CI 
\([0.584, 0.601]\)) than non-deliberating (median \(= 0.582\), 95\% CI \([0.573, 0.589]\)). The median paired 
difference was \(0.019\) (95\% CI \([0.006, 0.024]\)), with treatment showing greater mean reversion on 
45 of 71 items (63\%). While statistically significant (paired \(t = 4.05\), \(p < 0.001\), 
Cohen's \(d = 0.48\)), the absolute magnitude of the difference is small.

The MCF-2023 results show a somewhat stronger pattern. Treatment participants exhibited higher mean reversion (median \(= 0.564\), 95\% CI \([0.536, 0.586]\)) than controls (median \(= 0.526\), 95\% CI \([0.496, 0.542]\)). The median paired difference was \(0.048\) (95\% CI \([0.034, 0.062]\)), with treatment showing greater mean reversion on 38 of 44 items (86\%) (paired \(t = 8.93\), \(p < 0.001\), Cohen's \(d = 1.36\)).

Since some mean reversion is expected in any repeated measurement due to 
randomness, the relevant quantity is the \emph{excess} mean reversion in treatment 
relative to control. This excess (0.019 in MCF-2022; 0.048 in MCF-2023) suggests that deliberation may soften extreme 
opinions beyond what would occur through statistical artifact alone, though the effect sizes are modest and warrant cautious interpretation. 

Table~\ref{tab:opinion_dynamics} summarizes the findings from both MCF events. Across three dimensions of opinion change, deliberation is associated with patterns that differ from the control condition. Deliberation tends to generate greater opinion mixing and stronger mean reversion, with more heterogeneous variance effects. These patterns appear in both deliberation events despite differences in topics, languages, and participant populations, though effect sizes vary. Mean reversion effects are modest in MCF-2022 and somewhat stronger in MCF-2023. The next section complements these population-level findings with a discourse-level analysis linking conversational features to opinion change in the SOF dataset.

\subsection{Linking Discussion Features to Opinion Change}

The SOF dataset enables us to probe the \emph{mechanisms} driving opinion change by analyzing how specific features of small-group discourse relate to subsequent attitude shifts. We coded 6,232 participant statements across 460 room-agenda item pairs using an LLM that scores contributions on \emph{justification quality}, \emph{novelty}, and \emph{expressed support} (see \S\ref{sec:method} for details). Table~\ref{tab:sof_regression} presents the main results.

\begin{table}[h]
\centering
\small
\setlength{\tabcolsep}{2pt}  
\begin{tabular}{lcc}
\toprule
& \textbf{All Participants}
& \parbox[c]{2.5cm}{\centering \textbf{Non-contributing\\Participants Only}} \\
\midrule
Expressed Support & $1.603^{**}$ & $1.190^{**}$ \\
                 & (0.131) & (0.243) \\
Justification & $-0.034$ & $0.401^*$ \\
                     & (0.093) & (0.169) \\
Novelty & $0.032$ & $0.048$ \\
        & (0.146) & (0.363) \\
Support $\times$ Justification & $0.470^*$ & $0.913^*$ \\
                               & (0.203) & (0.394) \\
Support $\times$ Novelty & $-0.118$ & $-0.071$ \\
                        & (0.292) & (0.932) \\
Log(Num Statements) & $0.154$ & $-0.014$ \\
                & (0.116) & (0.167) \\
Pre-opinion (centered) & $-0.458^{**}$ & $-0.568^{**}$ \\
                      & (0.054) & (0.110) \\
\midrule
Agenda fixed effects & [-0.327, 0.989] & [-0.518, 1.163] \\
\midrule
$R^2$ & 0.534 & 0.212 \\
$N$ & 460 & 460 \\
\bottomrule
\end{tabular}

\vspace{1mm}
\raggedright
\footnotesize \textit{Note:} Robust standard errors (HC3) in parentheses. Agenda fixed effects included; range of coefficients shown (reference category omitted). $^{**}p<0.001$, $^*p<0.05$.
\caption{Regression predicting mean opinion change at the room-agenda level.}
\label{tab:sof_regression}
\end{table}
%
\begin{figure*}[h]
\centering
\begin{subfigure}{0.49\textwidth}
    \centering
    \includegraphics[width=\linewidth]{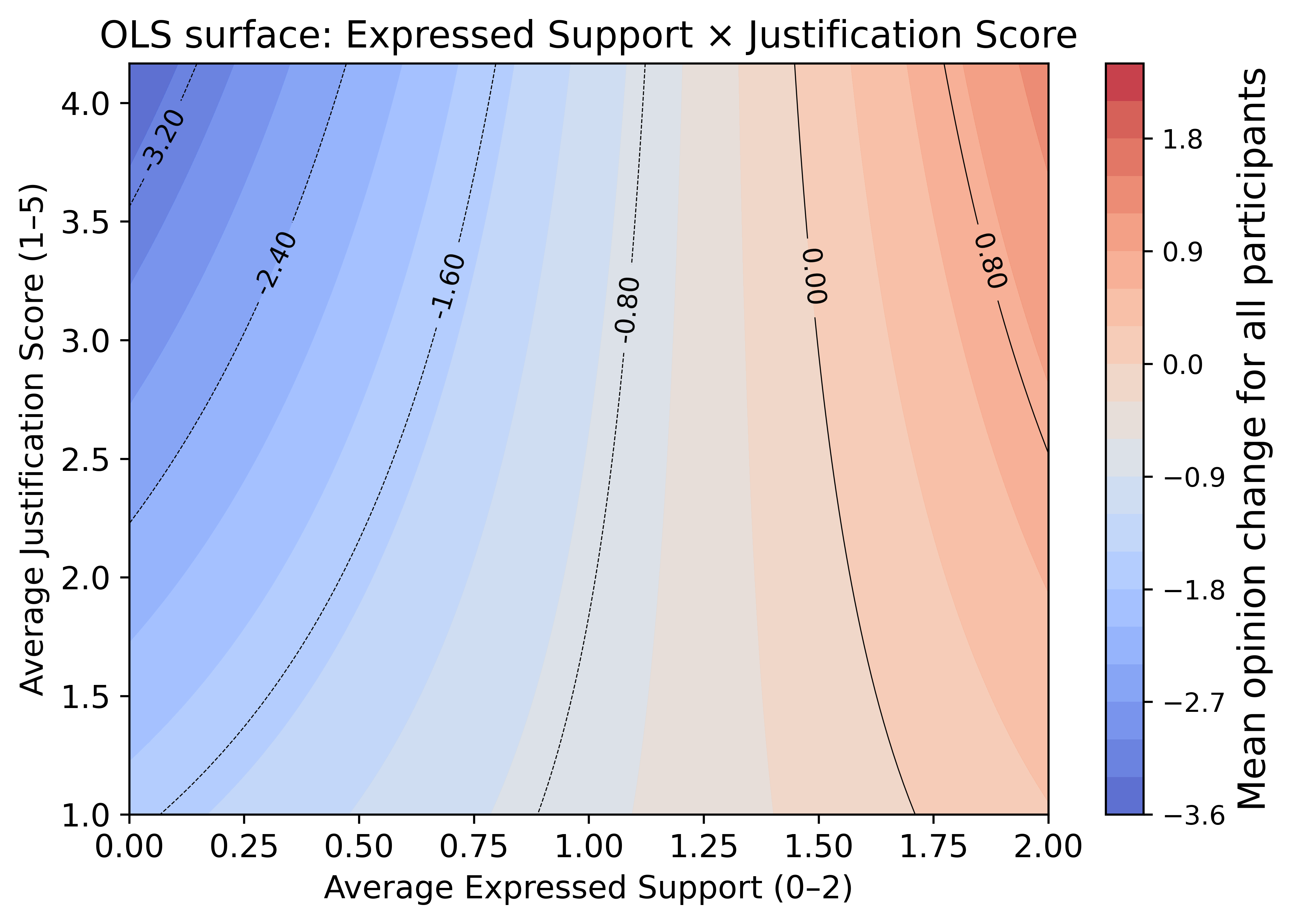}
    \caption{All participants. Higher justification quality is associated with a greater effect of expressed support on opinion change.}
    \label{fig:sof_interaction}
\end{subfigure}
\hfill
\begin{subfigure}{0.49\textwidth}
    \centering
    \includegraphics[width=\linewidth]{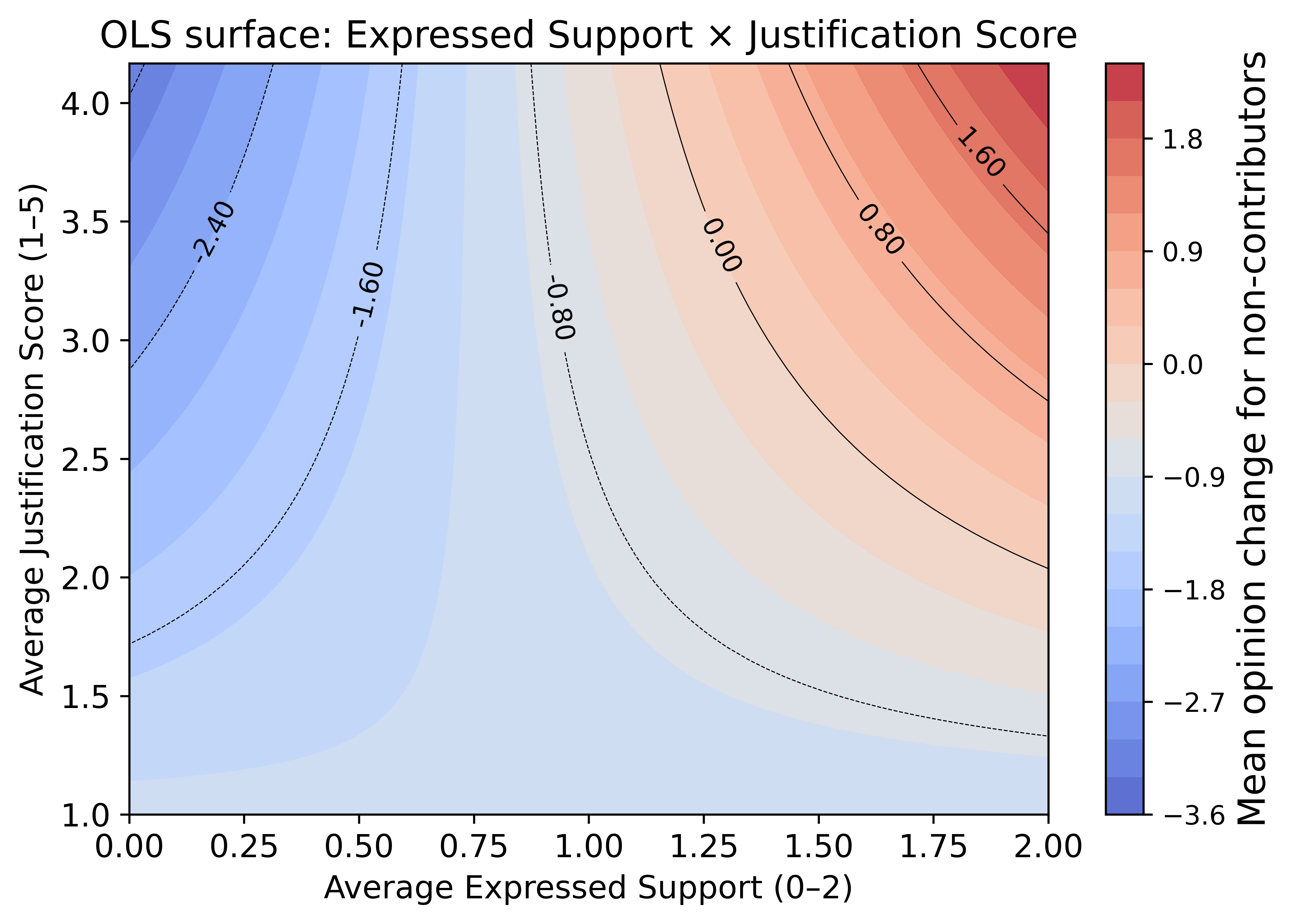}
    \caption{Non‑contributing participants only. The association of higher justification quality with a higher effect of expressed support persists.}
    \label{fig:sof_interaction_observers}
\end{subfigure}
\caption{Surface plots illustrating the interaction between expressed support and justification quality on opinion change. The left panel includes all participants; the right panel restricts to participants who did not contribute to the discussion on that topic.}
\label{fig:sof_interaction_combined}
\end{figure*}

\subsubsection{Expressed Support Is Associated With Opinion Change}
The strongest predictor of opinion change is the average level of expressed support in the discussion. A one-unit increase in mean support (on the 0--2 scale) is associated with a 1.60-point shift in average opinion (on the 0--10 scale) in the same direction ($p < 0.001$). This effect size is substantial: comparing a room where statements were on average neutral (support = 1) to one where they were more supportive (support = 1.5) predicts a  0.8-point higher increase for the latter room in post-deliberation agreement with the proposal.

Crucially, this association persists even for participants who did not speak during the discussion on that agenda item. Among the non-contributing participants (28.4\% of participants on average), a one-unit increase in expressed support predicts a 1.19-point opinion shift ($p < 0.001$). While attenuated compared to the full sample, this association demonstrates that exposure to supportive arguments is associated with change in attitudes even without active participation.

\subsubsection{Justification Quality Is Associated with Amplified Relation between Expressed Support and Opinion Change}
The moderation analysis reveals that the \emph{quality of justification} significantly moderates the relationship between expressed support and opinion change. The positive interaction coefficient (0.470, $p = 0.021$) indicates that well-justified supportive arguments are more persuasive than poorly justified ones. Figure~\ref{fig:sof_interaction} visualizes this interaction. This finding aligns with deliberative theory's emphasis on reason-giving as the mechanism of opinion change.

\subsubsection{No Significant Moderation is Observed with Novelty}
In contrast to justification quality, the novelty of arguments, as captured by our LLM-based annotations, shows no significant main effect or interaction with expressed support. The interaction coefficient is small and non-significant ($-0.118$, $p = 0.685$), suggesting that introducing new perspectives or information does not systematically enhance persuasion beyond the effect of supportive stance-taking. This null result, if robust, challenges information-centric models of deliberation and suggests that \emph{how} arguments are made matters more than whether they introduce novel content. However, as with most null results, this should be interpreted with caution. An absence of observed effect does not equate with an observation of absence of effect. It may just mean that we need to revisit our measure of novelty.

\subsubsection{Analysis on the Non-contributing Participants For Robustness}
The analysis of the participants who do not contribute any statement on a given agenda item is in Table~\ref{tab:sof_regression} (right column). It addresses potential endogeneity concerns; participants' expressed support could reflect rather than cause their opinion change. Among those who only listened, the pattern holds: expressed support predicts opinion change (1.19, $p < 0.001$), and this effect is amplified by justification quality (interaction = 0.913, $p = 0.021$). This impact of justification quality is even greater than that for the full population. Figure~\ref{fig:sof_interaction_observers} shows this moderation effect for non-contributors. While the reduced $R^2$ (0.212 vs.\ 0.534) indicates that active participation explains additional variance, the core mechanism of justified support influencing attitudes operates even through passive exposure.

\section{Discussion and Conclusion}
\label{sec:conclusion}

Our analysis of three large-scale deliberative polling events reveals consistent patterns in how structured deliberation reshapes participants' opinions. First, deliberation is associated with substantial opinion mixing: participants reorder their positions relative to one another rather than simply shifting together. The median Kendall's $\tau$ drops from 0.296 to 0.248 in MCF-2022 and from 0.385 to 0.327 in MCF-2023, with treatment groups showing lower rank correlation than controls on 97\% and 93\% of items, respectively. This finding, replicated across two events with different topics and participant populations, suggests that deliberation facilitates genuine reconsideration of positions rather than uniform drift.

Second, variance changes are heterogeneous and context-dependent. While control groups uniformly reduce variance (97\% of items in MCF-2022, 98\% in MCF-2023), treatment groups show both increases and decreases, with 39\% of items in MCF-2022 and 9\% in MCF-2023 exhibiting greater disagreement post-deliberation. Notably, deliberation amplifies the dispersion of variance effects: the standard deviation of variance changes across items is three times larger in treatment than control for MCF-2022 and 1.6 times larger for MCF-2023. This heterogeneity reconciles conflicting literatures on polarization versus consensus; deliberation can produce either outcome depending on the issue, the arguments raised, and the initial distribution of opinions.

Third, deliberation is associated with modestly greater mean reversion than control conditions. Treatment groups show higher mean reversion on 63\% of items in MCF-2022 and 86\% in MCF-2023. While the consistency across events suggests that exposure to diverse perspectives may soften extreme positions beyond what would occur through re-measurement alone, the absolute effect sizes are moderate.

Fourth, the SOF transcript analysis illuminates potential mechanisms underlying these patterns. Expressed support within discussions strongly predicts opinion change ($\beta = 1.60$, $p < 0.001$), and this effect is amplified when arguments include high-quality justifications (interaction $\beta = 0.47$, $p = 0.021$). No statistically significant moderating effect is observed with argument novelty. These patterns remain significant even among participants who only observed without speaking, suggesting that at least part of the mechanism includes exposure to well-reasoned discourse rather than through active participation itself.

Together, these findings have practical implications for deliberation design. The observed importance of justification quality over novelty suggests that facilitators should emphasize reason-giving (encouraging participants to support their positions with examples and evidence) rather than prioritizing the introduction of new information per se. The heterogeneity in variance outcomes also implies that deliberation's effects depend critically on contextual factors, cautioning against one-size-fits-all expectations.

Our results clarify both the scope and limits of deliberation's effects. The replication of opinion mixing and mean reversion patterns across MCF-2022 and MCF-2023 -- events covering distinct topic scopes, overlapping but different countries, and both aiming for representative samples of the populations in those countries -- strengthens confidence in the robustness of these phenomena. However, the modest magnitude of mean reversion effects and the context-dependent variance changes caution against universal claims about deliberation's moderating influence. Deliberation appears to work through the selective adoption of arguments based on their quality and alignment with prior beliefs, producing complex patterns of movement that vary by issue and population. Additional analysis on deliberations with different topics, moderation principles and participant characteristics could provide even more robust and additional insights. 

Several limitations warrant acknowledgment. The absence of a control group for SOF prevents us from isolating deliberation effects from panel conditioning in that dataset. Our analysis of non-contributing participants mitigates but does not eliminate endogeneity concerns in the mechanism analysis. The exclusion of ``no opinion'' responses may introduce selection effects if participants who form opinions during deliberation differ systematically from those with stable views. Finally, while our findings replicate across two MCF events, generalization to other deliberation formats and platforms remains an open question. Future work should examine whether these patterns hold in in-person deliberation, adversarial settings, or when participants have stronger prior commitments.

\paragraph{Ethical Considerations}
This research follows standard ethical guidelines for human subjects research. All participants provided informed consent through the \ifanonelse{\redacted}{Stanford} Online Deliberation Platform and were notified of data collection procedures for research purposes. Personally identifying information was removed before analysis, and all reported results are aggregated to prevent re-identification.

The use of large language models for discourse evaluation raises additional considerations. Although LLMs enable scalable analysis, their judgments may encode biases present in training data. We mitigate this concern through rubric-based prompts emphasizing reasoning quality and neutrality, and through validation against human annotations. Future work should continue to assess the fairness and transparency of LLM-based coding.

We utilized LLMs to assist with the preparation of this manuscript, including paraphrasing text, identifying related work, and generating code for data analysis.

\section*{Acknowledgments}
This study uses data that was collected under Institutional Review Board approval at Stanford University. Data collection and analysis were performed by researchers at that institution. Munagala contributed to study design discussions, performance metric development, and manuscript writing. Munagala was supported by the NSF grant IIS-2402823. 

We acknowledge the work by the Stanford Deliberative Democracy Lab in data collection and thank them for making survey data available. We thank Arjun Karanam for help with organizing and matching the SOF data. 


\bibliography{refs}


\newpage
\appendix

\section{Deliberation Agendas} \label{app:agendas}
In the SOF deliberation, the agenda items on Day 1 were:
    \begin{enumerate}
        \item  Elect the president by a national popular vote.
        \item Replace winner-take-all with fractional proportional.           
        \item Replace plurality voting with ranked-choice voting for the president. 
    \end{enumerate}

\noindent The following agenda items were discussed on Day 2. CCC stands for the Civilian Climate Corps.
    \begin{enumerate}
        \item The CCC Should Receive Enough Funding to have 3 Million Corps Members, Compensated With a Living Wage.                          
        \item The communities most vulnerable to climate change should determine the projects and priorities of the Civilian Climate Corps.
        \item Adopt a regional minimum wage that reflects differences in the cost of living across the U.S.                                  
        \item Give cash grants of \$1,000/month to all people over 18 years of age.                                                            
        \item Impose an annual wealth tax of 2\% on any wealth over \$50 million and 3\% for wealth over \$1 billion.
    \end{enumerate}

In MCF-2022, the agenda list was as follows. While these headings may be terse, the deliberation participants were able to see a more detailed description and common discussion points with each agenda item on the interface. 

    \begin{enumerate}
        \item Video capture.
        \item Video capture only in spaces where repeated bullying and harassment occur.
        \item Automatic speech detection.
        \item Bullying and harassment in members-only spaces.
        \item Speech detection alerts in public spaces.
        \item Speech detection alerts in members-only spaces.
        \item Visible moderators.
        \item Invisible moderators.
        \item Notify users when entering space with repeated bullying.
        \item  Visibility of members-only spaces
        \item Creator training and possible creator restrictions.
        
    \end{enumerate}

\section{Human Validation of Stance Annotations}
\label{app:stance_validation}

\subsection{Annotation Procedure}
We sampled 40 English statements (5 per agenda item) uniformly at random, excluding statements used as few-shot examples in the LLM prompt. Four graduate-student raters at a research university independently labeled each statement's stance toward the relevant agenda item on a three-point scale: 0 = oppose, 1 = neutral/unclear/mixed, 2 = support. Annotators spent approximately one minute per statement on average.

\subsection{Inter-Rater Reliability}
We report two inter-rater reliability (IRR) metrics:
\begin{itemize}[leftmargin=12pt]
    \item \textbf{Krippendorff's $\alpha$ (ordinal):} 0.558 (95\% bootstrap CI [0.370, 0.706]), which penalizes extreme disagreements (0 vs.\ 2) more than adjacent ones.
    \item \textbf{Fleiss' $\kappa$ (nominal):} 0.397 (95\% bootstrap CI [0.239, 0.541]), treating labels as unordered categories.
\end{itemize}
All four raters gave identical labels for 37.5\% of statements (15 of 40), while 65.0\% achieved at least three-of-four agreement (26 of 40). The higher ordinal than nominal reliability indicates that most disagreements were adjacent (neutral vs.\ support/oppose) rather than extreme. Bootstrap confidence intervals were computed over 10,000 resamples of the 40 statements.

\subsection{Leave-One-Out Evaluation of LLM Performance}
Because no objective ground truth exists for stance labels, we evaluate whether the LLM behaves like a typical human rater by comparing both to the human consensus. For each of the 160 statement--rater pairs (40 statements $\times$ 4 held-out raters), we computed the three-rater consensus (mean of the non-held-out raters) and recorded whether the LLM or the held-out human was closer to this target.

\paragraph{Pair-level results.} The LLM was closer to consensus in 36 cases (22.5\%), the held-out human in 10 cases (6.2\%), and 114 cases (71.2\%) were ties. This pattern held for all four raters individually: LLM wins exceeded human wins by margins of 12--2, 11--2, 7--4, and 6--2 across raters.

\paragraph{Statement-level results.} Aggregating across the four leave-one-out folds per statement, the LLM had lower average deviation from consensus on 21 statements, humans on 4 statements, and 15 were ties. Excluding ties, a two-sided binomial test rejects equal win rates ($p = 0.0009$, $n = 25$).

These results demonstrate that LLM stance labels track the central tendency of human judgments at least as well as an individual human rater, supporting scalable annotation.

    \section{Supporting Figures}\label{app:fig}
    
\begin{figure}[h]
    \centering
    \includegraphics[width=0.4\textwidth]{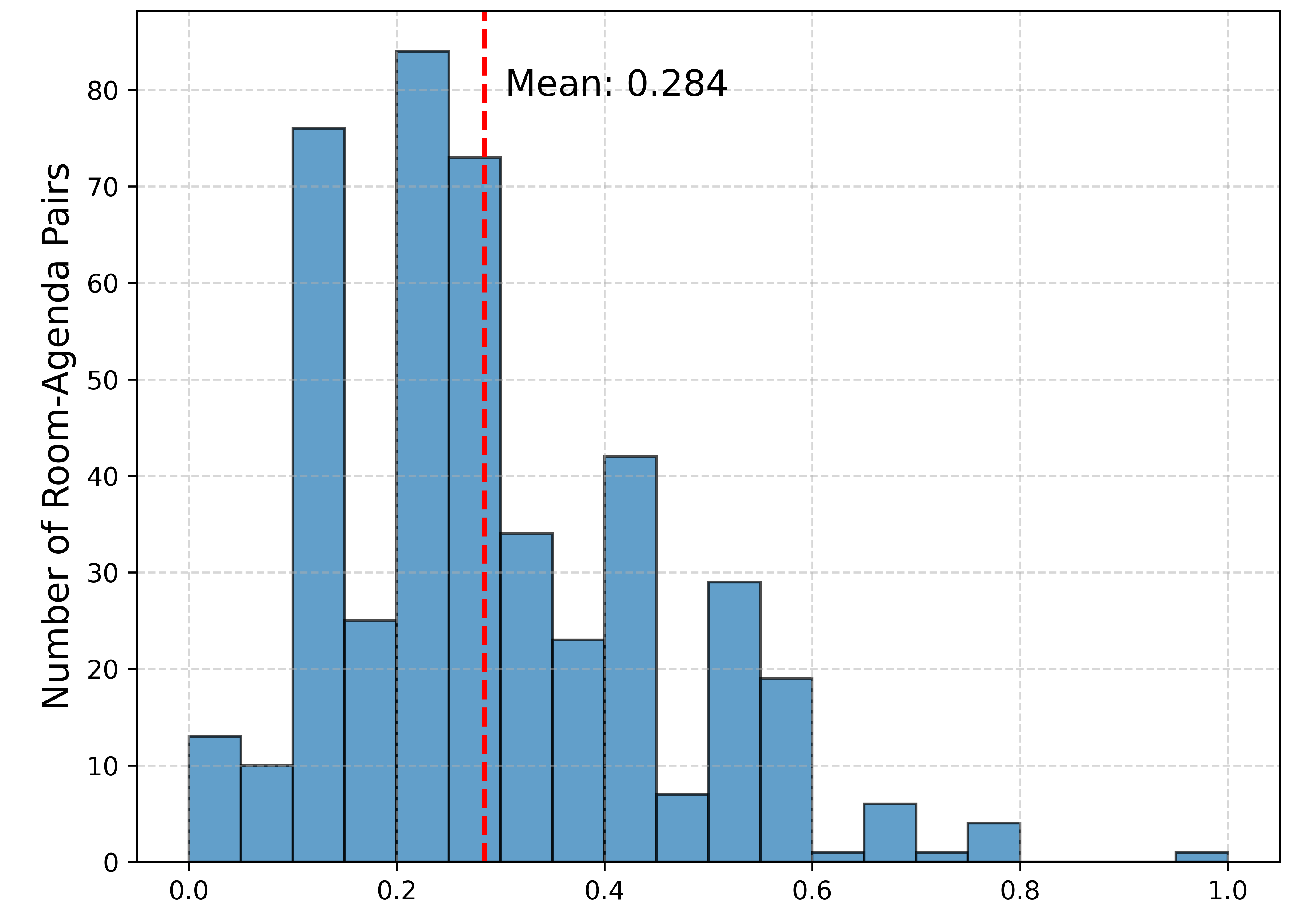} 
    \caption{Distribution of the fraction of non-contributing participants across room-agenda pairs. The red dashed line indicates the mean fraction (0.284).}
    \label{fig:observer_dist}
\end{figure}

\end{document}